\documentclass[aps,prb,reprint,superscriptaddress]{revtex4-2}
\usepackage[T1]{fontenc}
\usepackage[utf8]{inputenc}
\usepackage{graphicx}
\usepackage[svgnames]{xcolor}
\usepackage[colorlinks=true,linkcolor=RoyalBlue,citecolor=RoyalBlue,urlcolor=RoyalBlue]{hyperref}

\begin{document}

\title{Heat conduction in herbertsmithite: \\ field dependence at the onset of the quantum spin liquid regime}

\author{Quentin Barth\'elemy}
\email{quentin.barthelemy@usherbrooke.ca}
\affiliation{Institut Quantique, D\'epartement de physique \& RQMP, Universit\'e de Sherbrooke, Sherbrooke, Qu\'ebec J1K 2R1, Canada}

\author{\'Etienne Lefran\c cois}
\affiliation{Institut Quantique, D\'epartement de physique \& RQMP, Universit\'e de Sherbrooke, Sherbrooke, Qu\'ebec J1K 2R1, Canada}

\author{Jordan Baglo}
\affiliation{Institut Quantique, D\'epartement de physique \& RQMP, Universit\'e de Sherbrooke, Sherbrooke, Qu\'ebec J1K 2R1, Canada}

\author{Patrick Bourgeois-Hope}
\affiliation{Institut Quantique, D\'epartement de physique \& RQMP, Universit\'e de Sherbrooke, Sherbrooke, Qu\'ebec J1K 2R1, Canada}

\author{Dipranjan Chatterjee}
\affiliation{Universit\'e Paris-Saclay, CNRS, Laboratoire de Physique des Solides, 91405, Orsay, France}

\author{Pierre Leflo\"\i c}
\affiliation{Institut Quantique, D\'epartement de physique \& RQMP, Universit\'e de Sherbrooke, Sherbrooke, Qu\'ebec J1K 2R1, Canada}

\author{Matias Vel\'azquez}
\affiliation{Universit\'e Grenoble Alpes, CNRS, Grenoble INP, SIMAP, 38000, Grenoble, France}

\author{Victor Bal\'edent}
\affiliation{Universit\'e Paris-Saclay, CNRS, Laboratoire de Physique des Solides, 91405, Orsay, France}

\author{Bernard Bernu}
\affiliation{Sorbonne Universit\'e, CNRS, Laboratoire de Physique Th\'eorique de la Mati\`ere Condens\'ee, 75005, Paris, France}

\author{Nicolas Doiron-Leyraud}
\affiliation{Institut Quantique, D\'epartement de physique \& RQMP, Universit\'e de Sherbrooke, Sherbrooke, Qu\'ebec J1K 2R1, Canada}

\author{Fabrice Bert}
\affiliation{Universit\'e Paris-Saclay, CNRS, Laboratoire de Physique des Solides, 91405, Orsay, France}

\author{Philippe Mendels}
\affiliation{Universit\'e Paris-Saclay, CNRS, Laboratoire de Physique des Solides, 91405, Orsay, France}

\author{Louis Taillefer}
\email{louis.taillefer@usherbrooke.ca}
\affiliation{Institut Quantique, D\'epartement de physique \& RQMP, Universit\'e de Sherbrooke, Sherbrooke, Qu\'ebec J1K 2R1, Canada}
\affiliation{Canadian Institute for Advanced Research, Toronto, Ontario M5G 1M1, Canada}

\date{\today}

\begin{abstract}
We report thermal conductivity measurements on single crystals of herbertsmithite, over a wide range of temperatures ($0.05-120$~K) in magnetic fields up to $15$~T. We also report measurements of the thermal Hall effect, found to be vanishingly small. At high temperatures, in the paramagnetic regime, the thermal conductivity has a negligible field dependence. Upon cooling and the development of correlations, the onset of a clear monotonic field dependence below about $20$~K signals a new characteristic temperature scale that may reflect the subtle crossover towards the quantum spin liquid regime. Deconfined spinons, if present, are not detected and phonons, as the main carriers of heat, are strongly scattered by the intrinsic spin excitations and the magnetic defects. In view of the colossal fields required to affect the intrinsic spins, most of the field-induced evolution is attributed to the progressive polarization of some magnetic defects. By elaborating a phenomenological model, we extract the magnetization of these main scattering centers which does not resemble the Brillouin function for free spins $1/2$, requiring to go beyond the paradigm of isolated paramagnetic spins. Besides, the onset of a nonmonotonic field dependence below about $2$~K underlines the existence of another characteristic temperature scale, previously highlighted with other measurements, and sheds new light on the phase diagram of herbertsmithite down to the lowest temperatures.
\end{abstract}

\maketitle

\section{Introduction}

The search for a quantum spin liquid (QSL) in two dimensions has been an enduring problem since the resonating valence bond liquid was proposed as a precursor to high-temperature superconductivity in cuprates~\cite{Anderson1987,Baskaran1987,Wen2002,Lee2006,Savary2016,Broholm2020}. In theory, it is common to construct QSL states using the projected slave boson (alternatively slave fermion) approach for the $t-J$ model at half-filling, assuming the electron is a composite of a spinless bosonic holon and a chargeless spin-$1/2$ fermionic spinon~\cite{Baskaran1987,Wen2002,Lee2006,Savary2016,Lu2017}. Within this spin-charge separation picture, holons and spinons only interact through an emergent $U(1)$ or $Z_{2}$ gauge field. As elementary excitations of the QSL, pairs of spinons are formed coherently when pairs of spins are excluded from singlet formation. Deconfined spinons can move away from each other as long as the valence bond liquid can seep between them, through singlet reorganization. In case of hole doping of the QSL, the holon condensation is expected to trigger the conversion of spinon pairs into true Cooper pairs. A central question is therefore the existence of these mobile fermionic spinons in real two-dimensional QSL candidate materials~\cite{Yamashita2010,Watanabe2016,Doki2018,Bourgeois-Hope2019,Ni2019,Akazawa2020,LiN2020,Lefrancois2022}.

The mineral herbertsmithite ZnCu$_{3}$(OH)$_{6}$Cl$_{2}$ is an emblematic QSL candidate, being one of the best realizations of the nearest neighbor $S=1/2$ kagome Heisenberg antiferromagnet~\cite{Shores2005,Mendels2007,Mendels2016,Norman2016}. At the vertices of geometrically perfect kagome planes, magnetic Cu$^{2+}$ ($S=1/2$) ions interact through a strong antiferromagnetic coupling between nearest neighbors $J\sim190$~K, that far exceeds all other in-plane or interplane couplings~\cite{Jeschke2013}. The main perturbations to the pure Heisenberg model consist of (i) the presence of copper ions on about $20$~\% interplane zinc sites, likely producing extended magnetic defects and an effective dilution of the kagome lattice~\cite{Freedman2010,Smaha2020,Khuntia2020,Barthelemy2022}, and (ii) a finite out-of-plane Dzyaloshinskii-Moriya component $D_{\mathrm{z}}\sim0.06(2)J$ combined with possible symmetric exchange anisotropy~\cite{Zorko2008,ElShawish2010,Han2012}. In zero external magnetic field, herbertsmithite does not order or freeze down to the lowest temperatures reached in experiments ($20$~mK $\sim J/9500$)~\cite{Mendels2007,Mendels2016} and features a broad continuum of low-energy spin excitations, a characteristic of fractionalization~\cite{Han2012-bis,Han2016}.

Determining the ground state of the nearest neighbor $S=1/2$ kagome Heisenberg antiferromagnet remains a major issue for the condensed matter community. As reflected by the massive proliferation of low-lying energy levels in the exact spectra of small clusters~\cite{Lecheminant1997,Sindzingre2009}, numerical studies point to a collection of very different states with comparable energies including gapped QSL, gapless QSL or valence bond solids~\cite{Sachdev1992,Lecheminant1997,Waldtmann1998,Ran2007,Singh2007,Hermele2008,Sindzingre2009,Yan2011,Depenbrock2012,Clark2013,Iqbal2013,Iqbal2014,He2017,Liao2017,Lu2017,Chen2018,Hotta2018,Lee2018,Zhu2018,Hering2019,Jiang2019,Lauchli2019,Zhang2020}. Sticking to the fermionic spinon approach, it is now broadly believed that the most relevant candidates are the $U(1)[0,\pi]$ and $Z_{2}[0,\pi]$ ($\alpha$ or $\beta$) gapless QSL~\cite{Ran2007,Hermele2008,Clark2013,Iqbal2013,Iqbal2014,He2017,Lu2017,Zhu2018,Hering2019,Zhang2020}. They feature a spinon band structure with Dirac nodes or small circular Fermi surfaces at the Fermi level (one spinon per site), as opposed to the large Fermi surface proposal for the triangular QSL candidates~\cite{Motrunich2005}.

Originating from interplane copper ions, magnetic defects which may impact the physics of the kagome planes have been a recurring impediment to probing the pure kagome behavior. In the most straightforward measurements, their contribution is dominant at low temperature and low magnetic field, appearing as a Curie-like tail in susceptibility or Schottky-like term in specific heat~\cite{Helton2007,deVries2008,Han2012,Han2014,Khuntia2020,Barthelemy2022}. Recent specific heat measurements in high magnetic fields up to $34$~T allowed for the first time to separate the QSL behavior of the kagome planes from the Schottky-like anomaly~\cite{Barthelemy2022}. In a very unexpected way, an effective dilution of the kagome planes had to be introduced while the corresponding specific heat was found to be field-independent, from $34$~T to zero field, following a power law $C_{\mathrm{p}}^{\mathrm{kago}}(T\rightarrow0)\propto T^{\alpha}$ with a nontrivial exponent $\alpha\sim1.5$ in between the metal-like $\alpha=1$ (mean-field expectation for a spinon Fermi surface) and the graphene-like $\alpha=2$ (mean-field expectation for Dirac spinons). This gapless algebraic temperature dependence is typical of a critical QSL but the invariance with field seriously questions the presence of fermionic excitations. In this context, $^{17}$O NMR and $^{63,65}$Cu NQR appear as ideal probes of the kagome planes and they show the existence of two classes of copper sites: about $60$~\% with a fast relaxation lack a spin excitation gap, in line with the specific heat results, while $40$~\% display a slow relaxation, the interpretation of which is still debated~\cite{Olariu2008,Khuntia2020,Wang2021,Mendels2022}.

In herbertsmithite, it is also now well established that some field-induced instabilities occur at very low temperatures $T<2$~K $\sim J/95$, a temperature scale that is most likely determined by the deviations from the pure nearest neighbor Heisenberg model. From the $^{17}$O NMR perspective, the QSL behavior is robust down to at least $1$~K ($\sim J/190$) whatever the field intensity up to at least $12$~T~\cite{Jeong2011}. At sub-Kelvin temperatures however, the Gaussian-like line broadens and the spin-lattice relaxation rate drops dramatically along an activation law, which indicates the transition to a random frozen phase with gapped excitations. When constructing a $B$-$T$ phase diagram out of these NMR experiments, the transition between the QSL regime and the frozen phase extrapolates to a possible quantum critical point on the order of $1.55$~T in the $T\rightarrow0$ limit~\cite{Jeong2011}. Such a field-induced transition from a gapless to a gapped regime was similarly reported in the case of Zn-brochantite, another QSL candidate material~\cite{Gomilsek2017}. Besides, torque magnetometry~\cite{Asaba2014}, ac susceptibility~\cite{Helton2009PhD} and specific heat~\cite{Barthelemy2022} measurements revealed an additional (more vertical) in-field crossover extending even slightly above $1$~K. Therefore, in order to examine the kagome QSL behavior outside of the energy range in which perturbations have their most severe impact, the focus should be either on measuring herbertsmithite in zero field down to the lowest temperatures or in finite fields at temperatures above $2$~K.

In the present study, through the prism of thermal transport measurements on herbertsmithite single crystals, we first aim to determine whether the emergent excitations that originate in the field-independent gapless $C_{\mathrm{p}}^{\mathrm{kago}}$ give rise to notable spin-mediated heat conduction. Indeed, just as electrons would do in a metal, deconfined spinons may carry heat in addition to phonons, with distinct signatures expected at low temperature depending on the class of the QSL. Recently, Huang \emph{et al.}~\cite{Huang2021} and Murayama \emph{et al.}~\cite{Murayama2021} have started to explore this issue but concentrated on a small range of very low temperatures (respectively $0.05-0.8$~K and $0.1-0.5$~K) in magnetic fields up to $15$~T, always away from the QSL regime, within the frozen phase, except in zero field. In both of the aforementioned studies, the authors reported a scarcely field-dependent (if not field-independent) thermal conductivity, the absence of any residual term at $T=0$, and thus claimed the absence of mobile fermionic excitations (whether gapped or gapless) with the ability to delocalize over large portions of the sample. Nevertheless, the limited set of data calls for the present, more comprehensive study, first to check the results independently and then to investigate potential spinon-phonon decoupling and thermal Hall effect. Extending the measurements to higher temperatures reveals that the thermal conductivity is clearly sensitive to the magnetic field below about $20$~K. In our data, the absence of a residual term at $T=0$, of spinon-phonon decoupling and of thermal Hall effect, even though it does not rule out the presence of deconfined spinons, allow for the rigorous conclusion that phonons are the main carriers of heat. More significant is our observation of two distinct regimes of field dependence, monotonic above about $2$~K and nonmonotonic below, which highlight two characteristic low-temperature scales. Our comprehensive study of the field dependence, which we ascribe to a strong phonon scattering from the whole spin ensemble, opens a new avenue for characterizing the magnetic defects and helps in clarifying the phase diagram of herbertsmithite down to the lowest temperatures. 

\section{Experimental details}

We performed thermal conductivity ($\kappa_{\mathrm{xx}}$) and thermal Hall conductivity ($\kappa_{\mathrm{xy}}$) measurements on high quality single crystals of protonated (the PHS sample) and deuterated (the DHS sample) herbertsmithite that were prepared using methods described elsewhere~\cite{Han2011,Velazquez2020}. We use the same labels as in Ref.~\cite{Barthelemy2022} because the samples are obtained in the same way and have similar copper contents. Based on SQUID susceptibility and specific heat data, the DHS sample harbors slightly more magnetic defects than the PHS sample~\cite{Barthelemy2022}. This is due to the different temperatures used in the growth process. Attributing the Curie-like tail in susceptibility and the Schottky-like term in specific heat to copper ions sitting on the interplane zinc sites leads to an occupation as high as $18$~\% for the PHS sample and $23$~\% for the DHS sample. The samples were measured as grown (uncut) for a wide range of temperatures ($0.05-120$~K) in magnetic fields up to $15$~T, working in variable temperature insert (VTI) and dilution refrigerator (DR) environments with thermocouple and ruthenium oxide temperature sensors respectively. The samples were first examined with X-ray diffraction to determine the orientation of the high symmetry axes. The largest faces, on which the placement of contacts is easier, were not found parallel to the kagome planes contrary to what was assumed in other studies~\cite{Huang2021,Murayama2021}.

The measurements were carried out by employing a standard steady-state method. A constant heat current $\mathbf{j}$ is injected at one end of the sample while the other end is thermally sunk to a copper block at bath temperature $T_{0}$. The heat current is generated when applying an electrical current through a strain gauge whose resistance marginaly depends on the temperature and magnetic field (about $5$~k$\Omega$ in the VTI and about $10$~k$\Omega$ in the DR). Assuming a one-dimensional heat flow and an isotropic medium, the longitudinal thermal gradient $\mathrm{d}T_{\mathrm{x}}$ is measured between two contacts separated by a distance $l$ along the heat flow. This gradient is evaluated using absolute and differential type-E thermocouples in the VTI, and ruthenium oxide sensors in the DR (calibrated \emph{in situ} against a reference germanium
thermometer placed in a field compensated zone). The longitudinal thermal conductivity is then given by $\kappa_{\mathrm{xx}}=j/(\alpha\mathrm{d}T_{\mathrm{x}})$ where $\alpha$ is a geometrical factor determined by the cross section $wt$ ($w$: width, $t$: thickness) divided by $l$. In the VTI, in presence of a magnetic field $\mathbf{B}\perp\mathbf{j}$, the transverse thermal gradient $\mathrm{d}T_{\mathrm{y}}$ is measured between two contacts separated by the distance $w$ perpendicular to the heat flow. This gradient is evaluated using a differential type-E thermocouple. After antisymmetrization, $\mathrm{d}T_{\mathrm{y}}=[\mathrm{d}T_{\mathrm{y}}(\mathbf{B})-\mathrm{d}T_{\mathrm{y}}(-\mathbf{B})]/2$, the thermal Hall conductivity is given by $\kappa_{\mathrm{xy}}=l\kappa_{\mathrm{yy}}\mathrm{d}T_{\mathrm{y}}/(w\mathrm{d}T_{\mathrm{x}})=l\kappa_{\mathrm{xx}}\mathrm{d}T_{\mathrm{y}}/(w\mathrm{d}T_{\mathrm{x}})$.

The contacts consisted of 17~$\mu$m gold wires attached with silver paint and had geometries ($l\times w\times t$): $526\times774\times326$~$\mu$m$^{3}$ for the PHS sample in the DR, $609\times293\times266$~$\mu$m$^{3}$ for the DHS sample in the DR and $644\times282\times334$~$\mu$m$^{3}$ for the (same) DHS sample in the VTI. Uncertainties on the absolute values of $\kappa_{\mathrm{xx}}$, on the order of $20$~\% at maximum, mainly originate from the uncertainties on the dimensions $l$, $w$ and $t$. The heat current $\mathbf{j}$ was chosen such that $\mathrm{d}T_{\mathrm{x}}/T\sim5$-$10$~\% and the resulting $\kappa_{\mathrm{xx}}$ was independent of $\mathrm{d}T_{\mathrm{x}}/T$, indicating no heat loss. Although we made sure that $\mathbf{j}$ had a large in-plane component $j_{\mathrm{ab}}$, let us recall that the magnetic anisotropy is weak at low temperature~\cite{Han2012} and that spinon transport (if any) is expected to occur in all directions, even between the planes~\cite{Werman2018}. Therefore, the orientations of $\mathbf{j}$ and $\mathbf{B}$ do not really matter for the present study. We always applied the magnetic field perpendicular to $\mathbf{j}$ and we did not notice any major difference in terms of shape and magnitude between the data obtained in our three configurations and elsewhere~\cite{Huang2021,Murayama2021}. The small mismatch observed in the DHS sample’s $\kappa_{\mathrm{xx}}$ between DR and VTI data (enhanced in the semilog display of Fig.~\ref{fig:Publi-Figure-1}) mostly arises from the error made when estimating the dimensions $l$, $w$ and $t$.

\section{Results}

\subsection{\label{sec:A}From the paramagnetic to the QSL regime}

\begin{figure}[t!]\centering
\includegraphics[width=\linewidth]{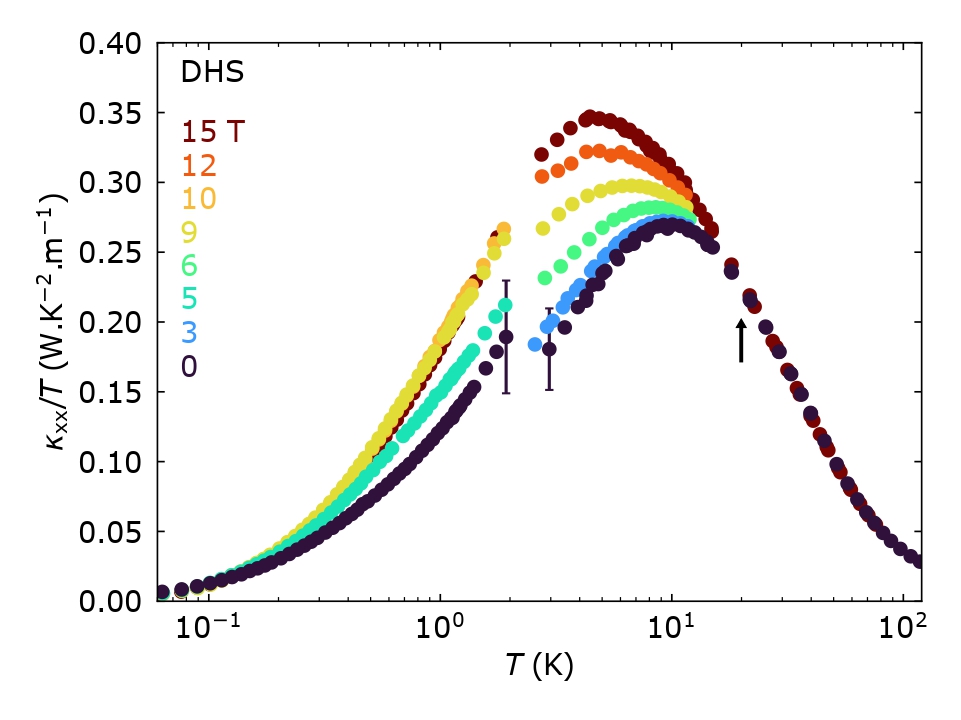}
\caption{\label{fig:Publi-Figure-1}Thermal conductivity of the DHS sample, as a semilog plot of $\kappa_{\mathrm{xx}}/T$ vs $T$, for various magnetic fields applied perpendicular to the heat current. Below about $20$~K (arrow), $\kappa_{\mathrm{xx}}$ acquires a clear field dependence. We show typical error bars on both sides of the minor discontinuity at about $2$~K between the DR and VTI data.}
\end{figure}

\begin{figure}[t!]\centering
\includegraphics[width=\linewidth]{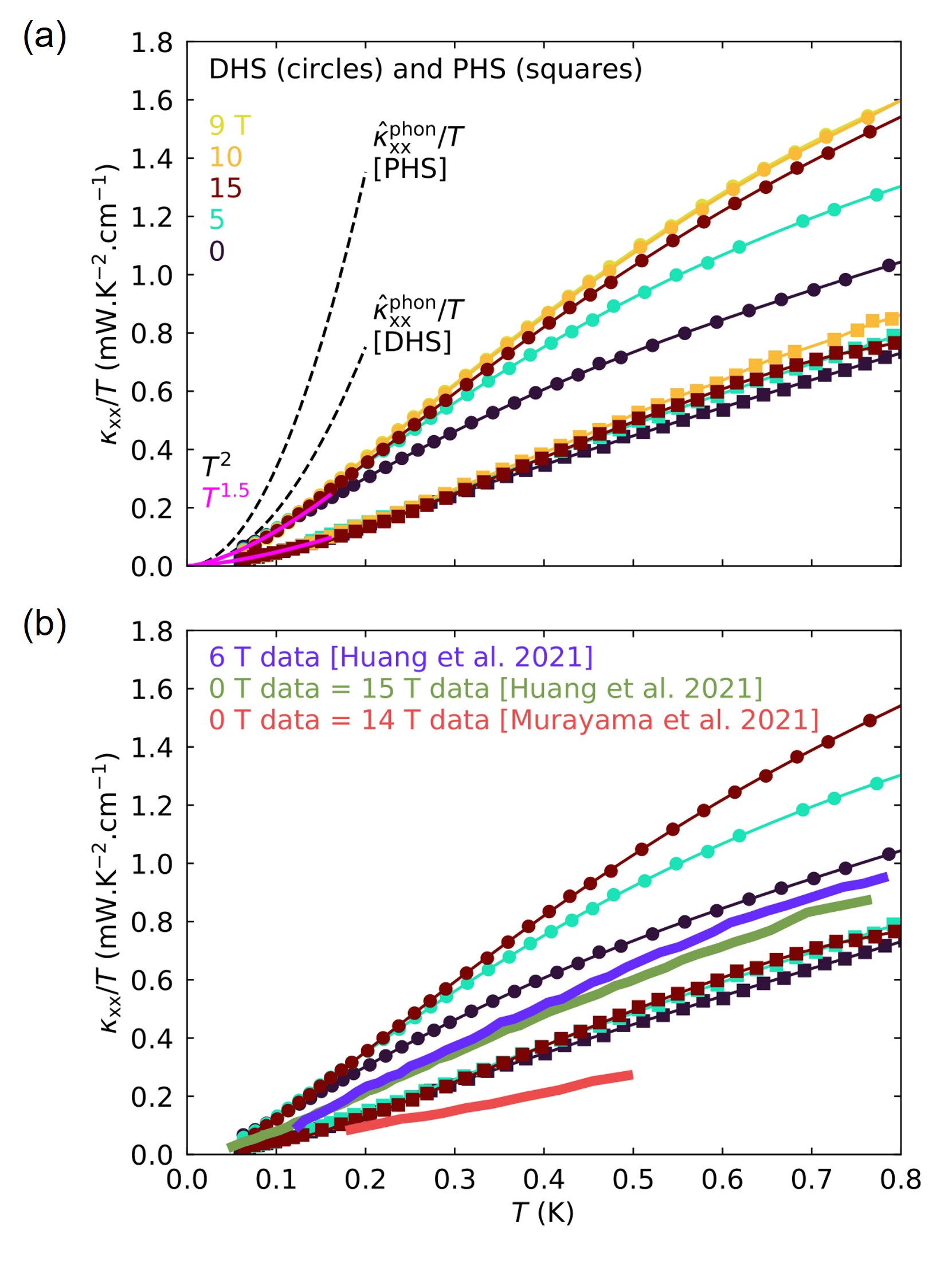}
\caption{\label{fig:Publi-Figure-2}{(a) Thermal conductivity of the two DHS and PHS samples at very low temperatures, as a plot of $\kappa_{\mathrm{xx}}/T$ vs $T$, for various magnetic fields applied perpendicular to the heat current. Solid lines are a guide to the eye. Below $160$~mK and whatever the field intensity, $\kappa_{\mathrm{xx}}/T\propto T^{1.5}$ for both samples (extrapolations in pink). Dashed lines depict estimates of $\hat{\kappa}_{\mathrm{xx}}^{\mathrm{phon}}/T$ in the boundary scattering limit. (b) Some data adapted from Refs.~\cite{Huang2021,Murayama2021} (thick lines) are superimposed on our data. The color code is the same in (a) and (b).}}
\end{figure}

\begin{figure}[t!]\centering
\includegraphics[width=\linewidth]{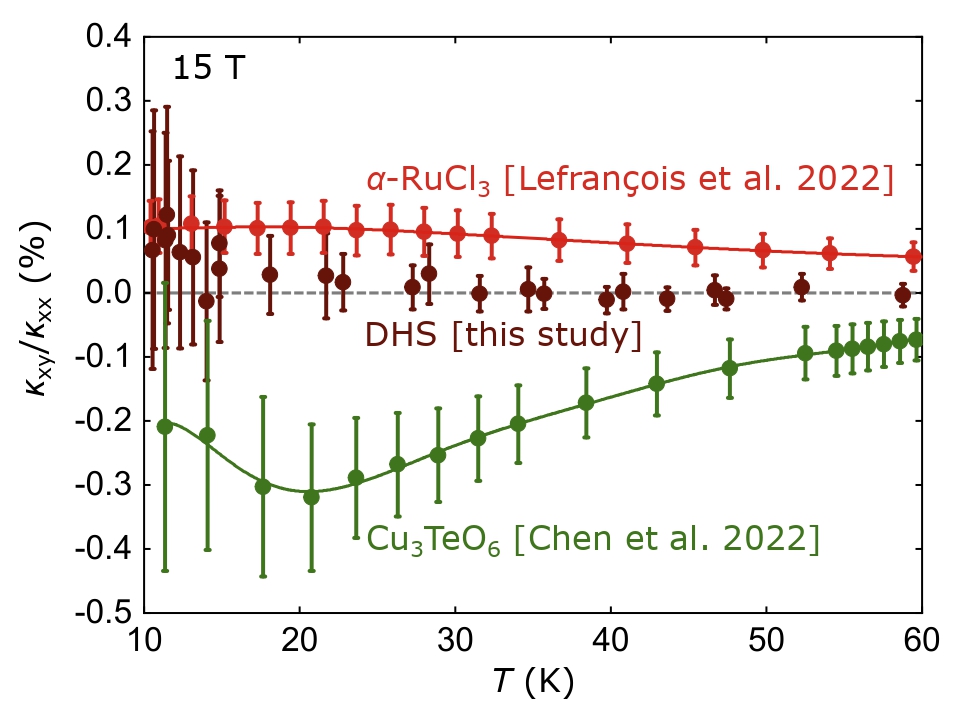}
\caption{\label{fig:Publi-Figure-3}A plot of the ratio $\kappa_{\mathrm{xy}}/\kappa_{\mathrm{xx}}$ vs $T$, for the DHS sample, $\alpha$-RuCl$_{3}$ (adapted from Ref.~\cite{Lefrancois2022}) and Cu$_{3}$TeO$_{6}$ (adapted from Ref.~\cite{Chen2022}), in a magnetic field of $15$~T applied perpendicular to the heat current. Solid lines are a guide to the eye. For the DHS sample, the ratio $\kappa_{\mathrm{xy}}/\kappa_{\mathrm{xx}}$ remains close to zero (dashed grey line) over a large temperature range $10-60$~K.}
\end{figure}

Let us focus first on the data representing the evolution from the high temperature paramagnetic regime to the QSL regime. We thus rely on the data obtained in zero field at all temperatures and the data obtained in finite fields at temperatures higher than $2$~K. These data show a smooth temperature evolution, without any anomaly that could indicate a sharp structural or magnetic transition. A striking feature however is the onset of a clear field dependence below about $20$~K, see Fig.~\ref{fig:Publi-Figure-1}. This observation is a direct rebuttal to recent reports of a scarcely field-dependent (if not field-independent) thermal conductivity, which were based on very low temperature data only (below $0.8$~K)~\cite{Huang2021,Murayama2021}. Here we examine the origin of both the temperature and field dependences in the QSL regime, investigating the nature of the carriers of heat and their scatterers.

To date, the accumulation of numerical studies on the nearest neighbor $S=1/2$ kagome Heisenberg antiferromagnet strongly supports the stabilization of a Dirac or small Fermi surface QSL ground state with mobile fermionic spinons~\cite{Ran2007,Hermele2008,Clark2013,Iqbal2013,Iqbal2014,He2017,Lu2017,Zhu2018,Hering2019,Zhang2020}. In zero field, when embedding some defects, both of these scenarios yield a small but finite density of states at zero temperature~\cite{Alloul2009,Werman2018}. In principle, this should result in a residual term in the thermal conductivity [a finite $\kappa_{\mathrm{xx}}(T\rightarrow0,B=0)/T$]. In practice, the accurate detection of such a residual term is challenging and may be prevented by a possible thermal decoupling of spinons from phonons at the lowest temperatures. As it was evidenced with several measurements on some high-$T_{\mathrm{c}}$ cuprates in both the normal and superconducting states, a drop of the heat transfer rate from phonons (the primary heat providers at the hot end of the sample) to electrons (either Fermi sea normal electrons or $d$-wave Dirac quasiparticles) results in a steep downturn of the thermal conductivity, following $\kappa_{\mathrm{xx}}(T\rightarrow0)\propto T^{\eta}$ with $\eta$ between $4$ and $5$~\cite{Smith2005}. 

As verified over a large range of temperatures, see Fig.~\ref{fig:Publi-Figure-1}, our thermal conductivity data do not show any anomalous downturn so that we can confidently rule out any spinon-phonon decoupling. For both the DHS and PHS samples in zero field, our data can readily be seen to extrapolate to zero as $T\rightarrow0$, see Fig.~\ref{fig:Publi-Figure-2}, in good agreement with recent reports~\cite{Huang2021,Murayama2021}. This casts serious doubt on the presence of deconfined spinons in the ground state of herbertsmithite. Below $0.16$~K, for both samples, the best extrapolations correspond to power laws of the form $\kappa_{\mathrm{xx}}(T\rightarrow0,B=0)\propto T^{\eta}$ with $\eta\sim2.5$.

For temperatures from $10$ to $60$~K, we also report the absence of any noticeable transverse thermal gradient in the DHS sample when a magnetic field of $15$~T (the highest field intensity in our experiments) is applied perpendicular to the heat current. On this range, $|\mathrm{d}T_{\mathrm{y}}|$ is found always smaller than $0.5$~mK. We cannot rely on measurements at lower temperatures because the thermocouple sensitivity gets significantly degraded and is no longer sufficient to detect such small signals. The ``degree of chirality'', given by the ratio $|\kappa_{\mathrm{xy}}/\kappa_{\mathrm{xx}}|$, is much lower than the typical value on the order of $0.1$~\% measured in other insulators for which there is a small but substantial thermal Hall effect, like $\alpha$-RuCl$_{3}$ ($0.1$~\% at $20$~K) or Cu$_{3}$TeO$_{6}$ ($0.3$~\% at $20$~K), see Fig.~\ref{fig:Publi-Figure-3}. This reflects the absence of the large Lorentz-force-driven thermal Hall effect predicted for QSL states with fermionic spinons~\cite{Katsura2010}.

Both the lack of spinon-phonon decoupling and sizeable thermal Hall effect in the QSL regime were the missing elements in previous reports to state that there is no evidence of a significant contribution from mobile spin excitations to the heat conduction in herbertsmithite. If the emergent excitations are really mobile, it is then necessary for their mean free path to be drastically altered, so that they cannot carry the heat efficiently along the entire length of the sample. This would mean that they are either heavily scattered and/or compelled to delocalize over small bounded areas or closed paths like segments or loops. Conservatively, as already suggested by the field-independent specific heat of the kagome planes $C_{\mathrm{p}}^{\mathrm{kago}}$, we suspect that herbertsmithite does not harbor any neutral fermion. Still, thermal conductivity is not the probe to provide a definitive proof for this. It can only tell whether there is a sizeable spinon contribution on top of the phonon contribution, and this is not the case here. Hence, in the present study, we do not prove the absence of mobile spin excitations, which could ultimately challenge the establishment of a QSL in herbertsmithite, but rather provide evidence that phonons are the main carriers of heat.

\begin{figure}[t!]\centering
\includegraphics[width=\linewidth]{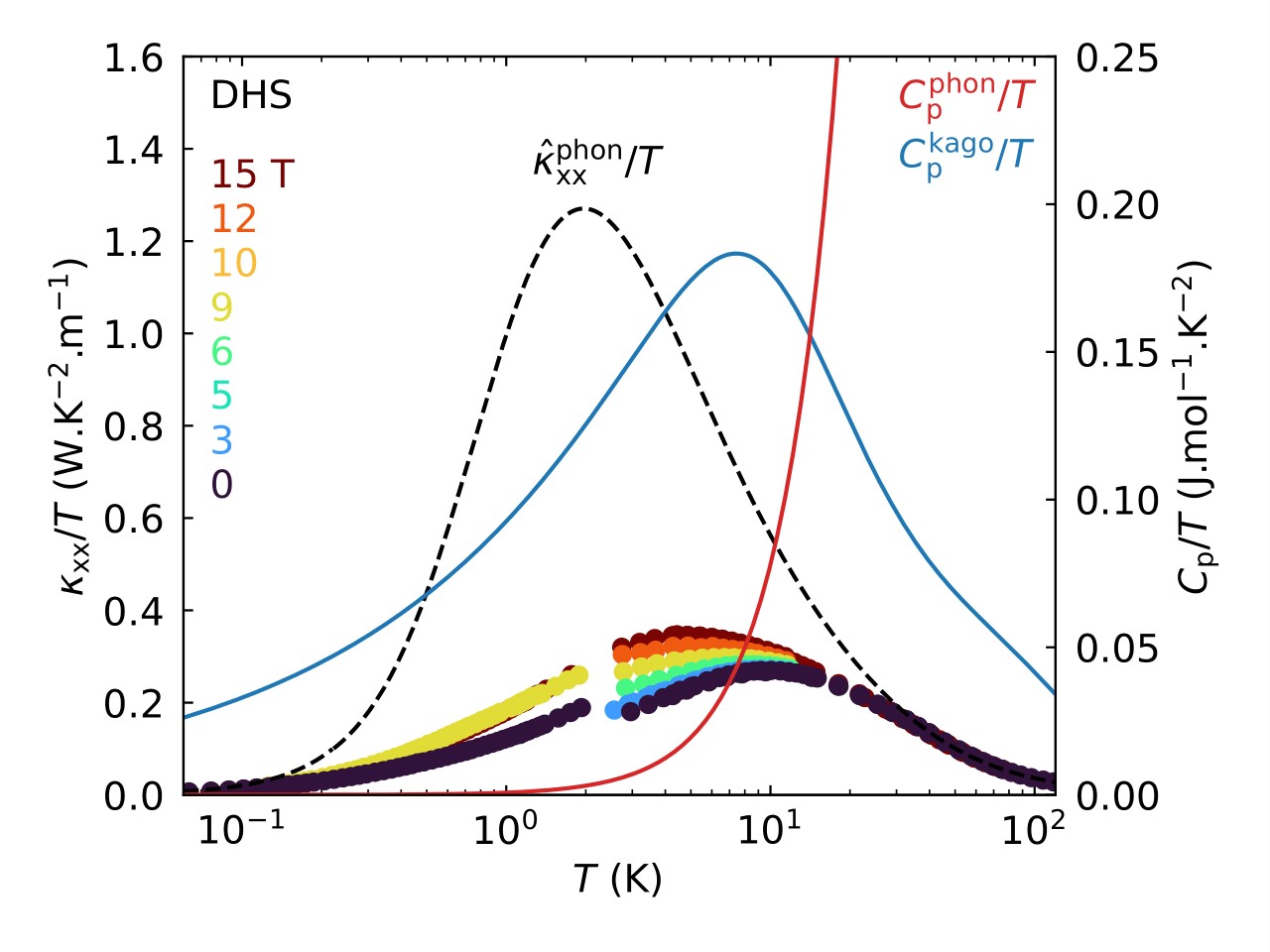}
\caption{\label{fig:Publi-Figure-4}Left axis: same data as in Fig.~\ref{fig:Publi-Figure-1} and corresponding estimate of $\hat{\kappa}_{\mathrm{xx}}^{\mathrm{phon}}/T$ over the whole temperature range (dashed line). Right axis: estimate of the phonon specific heat $C_{\mathrm{p}}^{\mathrm{phon}}/T$ and computed field-independent specific heat of the kagome planes $C_{\mathrm{p}}^{\mathrm{kago}}/T$ (adapted from Ref.~\cite{Barthelemy2022}).}
\end{figure}

Now, if the thermal transport is only ascribed to phonons, we must invoke a strong coupling to the spin degrees of freedom in order to explain the deviation from the expected $\eta=3$ to the observed $\eta\sim2.5$ along with the monotonic field dependence. In principle, at very low temperatures, ballistic phonons are primarily scattered by the sample boundaries. In this limit, the thermal conductivity is given by ${3\hat{\kappa}_{\mathrm{xx}}^{\mathrm{phon}}=C_{\mathrm{p}}^{\mathrm{phon}}v^{\mathrm{phon}}l^{\mathrm{phon}}}$. ${C_{\mathrm{p}}^{\mathrm{phon}}(T)=\beta T^{3}}$ is the phonon specific heat, ${v^{\mathrm{phon}}=(2\pi^{2}k_{\mathrm{B}}^{4}V_{\mathrm{m}}/5\beta\hbar^{3})^{1/3}}$ is the speed of sound where $V_{\mathrm{m}}\simeq114.3$~cm$^{3}$.mol$^{-1}$ is the molar volume of herbertsmithite and ${l^{\mathrm{phon}}=2(wt/\pi)^{1/2}}$ is the phonon mean free path set by the sample's width $w$ and thickness $t$. This mean free path is about $567$~$\mu$m for the PHS sample in the DR, $315$~$\mu$m for the DHS sample in the DR and about $346$~$\mu$m for the (same) DHS sample in the VTI. As specific heat measurements allowed us to determine $\beta=0.78(3)$~mJ.mol$^{-1}$.K$^{-4}$ (the corresponding Debye temperature $\theta_{\mathrm{D}}$ is about $151$~K)~\cite{Barthelemy2022}, see Fig.~\ref{fig:Publi-Figure-4}, we deduce $v^{\mathrm{phon}}\simeq2620$~m.s$^{-1}$ and compute $\hat{\kappa}_{\mathrm{xx}}^{\mathrm{phon}}$ for each sample, see Fig.~\ref{fig:Publi-Figure-2}. Actually, with the use of the Debye-Callaway model, we can even determine the complete evolution of $\hat{\kappa}_{\mathrm{xx}}^{\mathrm{phon}}$ from room temperature down to the lowest temperatures in the absence of any magnetic scattering, by fitting the defect and Umklapp scattering terms on the data above $40$~K:
\begin{equation}
\hat{\kappa}_{\mathrm{xx}}^{\mathrm{phon}}(T)=\frac{k_{\mathrm{B}}^{4}T^{3}}{2\pi^{2}\hbar^{3}v^{\mathrm{phon}}}\int_{0}^{\theta_{\mathrm{D}}/T}\frac{x^{4}e^{x}}{(e^{x}-1)^{2}}\tau_{0}(x,T)\mathrm{d}x,
\end{equation}
with $x=h\nu/(k_{\mathrm{B}}T)$ ($\nu$ is the phonon frequency) and the nonmagnetic scattering rate given by:
\begin{equation}
\frac{1}{\tau_{0}}=\frac{v^{\mathrm{phon}}}{l^{\mathrm{phon}}}+a\left(\frac{k_{\mathrm{B}}Tx}{\hbar}\right)^{4}+bT\left(\frac{k_{\mathrm{B}}Tx}{\hbar}\right)^{2}e^{-\theta_{\mathrm{D}}/3T},
\end{equation}
where $a$ and $b$ are the two fit parameters corresponding to the strengths of the defect and Umklapp scattering processes~\cite{Callaway1959,Slack1964}. In Fig.~\ref{fig:Publi-Figure-4}, we plot our best estimate for the DHS sample in the VTI, with $a=8(2)\times10^{-41}$~s$^{3}$ and $b=2(1)\times10^{-18}$~s.K$^{-1}$. We highlight that the measured thermal conductivity is about five times lower at $4$~K in zero field. This suggests, as in the case of many other QSL candidates~\cite{Fauque2015,Xu2016,Yu2018,Bourgeois-Hope2019}, that phonons are strongly scattered by (i) spin excitations of the quantum paramagnet (whether localized or mobile) and (ii) magnetic defects. Indeed, let us emphasize that $C_{\mathrm{p}}^{\mathrm{kago}}/T$, the Schottky-like anomaly and $\hat{\kappa}_{\mathrm{xx}}^{\mathrm{phon}}/T$ have a maximum within the same temperature window below $10$~K, in a range where the estimate for the phonon specific heat $C_{\mathrm{p}}^{\mathrm{phon}}/T$ drops rapidly ($C_{\mathrm{p}}^{\mathrm{phon}}/C_{\mathrm{p}}^{\mathrm{kago}}\sim8$~\% at $4$~K)~\cite{Barthelemy2022}, see Fig.~\ref{fig:Publi-Figure-4}.

The strong spin phonon coupling is also reflected in the evolution of the thermal conductivity with applied field, namely the monotonic field dependence which appears below about $20$~K down to about $2$~K, see Fig.~\ref{fig:Publi-Figure-1}. The gradual enhancement of $\kappa_{\mathrm{xx}}$ when increasing the field intensity is then understood in terms of a weakening of the magnetic scattering processes, when some spin fluctuations are suppressed due to the polarization of both the intrinsic spins in the kagome planes and the magnetic defects. Such a mechanism has already been proposed in the case of the Kondo insulator SmB$_{6}$, a diluted terbium pyrochlore oxide and the layered honeycomb magnet CrCl$_{3}$ for instance~\cite{Berman1976,Boulanger2018,Hirokane2019,Pocs2020}. Following Ref.~\cite{Pocs2020}, we can therefore consider a simple phenomenological model to describe the data, where the ratio $\hat{\kappa}_{\mathrm{xx}}^{\mathrm{phon}}(T)/\kappa_{\mathrm{xx}}(T,B)$ is expressed as $1+\xi(T,B)$ and the total scattering rate $1/\tau$ is given by the sum of the nonmagnetic scattering rate $1/\tau_{0}$ and a magnetic scattering rate $1/\tau_{\mathrm{m}}=\xi/\tau_{0}$. We can write $\xi(T,B)$ as $\lambda^{\mathrm{kago}}(T)n^{\mathrm{kago}}(T,B)+\lambda^{\mathrm{def}}(T)n^{\mathrm{def}}(T,B)$ where the field-independent $\lambda$ parameters represent effective strengths for the spin phonon coupling and the field-dependent $n$ parameters account for the fractions of non polarized spins, in both the kagome and magnetic defect systems. We have introduced two separate temperature-dependent couplings because these two systems are characterized by fluctuations with different frequency scales. In zero field, $n^{\mathrm{kago}}=n^{\mathrm{def}}=1$ so that $\kappa_{\mathrm{xx}}(T,B=0)=\hat{\kappa}_{\mathrm{xx}}^{\mathrm{phon}}(T)/[1+\lambda^{\mathrm{kago}}(T)+\lambda^{\mathrm{def}}(T)]$. When all the spins are fully polarized, which is never the case even in the strongest magnetic fields available~\cite{Okuma2020}, $n^{\mathrm{kago}}=n^{\mathrm{def}}=0$ so that $\kappa_{\mathrm{xx}}(T)$ reaches $\hat{\kappa}_{\mathrm{xx}}^{\mathrm{phon}}(T)$. Introducing the reduced magnetizations $m^{\mathrm{kago}}(T,B)=1-n^{\mathrm{kago}}(T,B)$ and $m^{\mathrm{def}}(T,B)=1-n^{\mathrm{def}}(T,B)$, the ratio of scattering rates is rewritten as $-\xi(T,B)=\lambda^{\mathrm{kago}}(T)m^{\mathrm{kago}}(T,B)+\lambda^{\mathrm{def}}(T)m^{\mathrm{def}}(T,B)-[\lambda^{\mathrm{kago}}(T)+\lambda^{\mathrm{def}}(T)]$.

\begin{figure}[t!]\centering
\includegraphics[width=\linewidth]{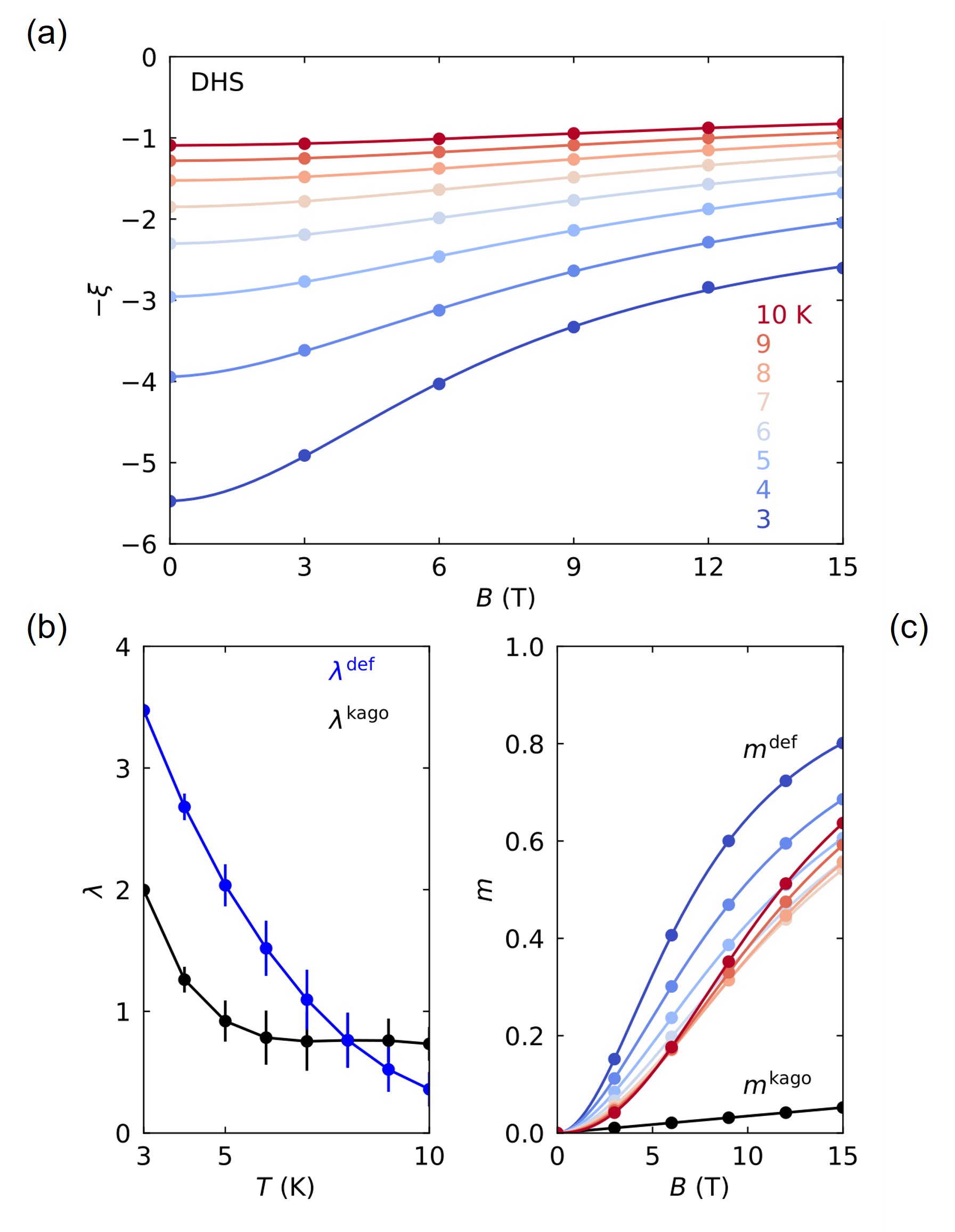}
\caption{\label{fig:Publi-Figure-5}(a) Field evolution of $-\xi$, extracted for several temperatures between $3$ and $10$~K. Solid lines are fitted curves, see text. (b) Temperature evolution of the fit parameters $\lambda^{\mathrm{def}}$ and $\lambda^{\mathrm{kago}}$. (c) Fitted magnetization profiles $m^{\mathrm{def}}$ vs $B$ for several temperatures between $3$ and $10$~K and fixed magnetization profile $m^{\mathrm{kago}}$ vs $B$ (based on the results of Ref.~\cite{Barthelemy2022}). The color scale is the same in (a) and (c).}
\end{figure}

\begin{figure}[t!]\centering
\includegraphics[width=\linewidth]{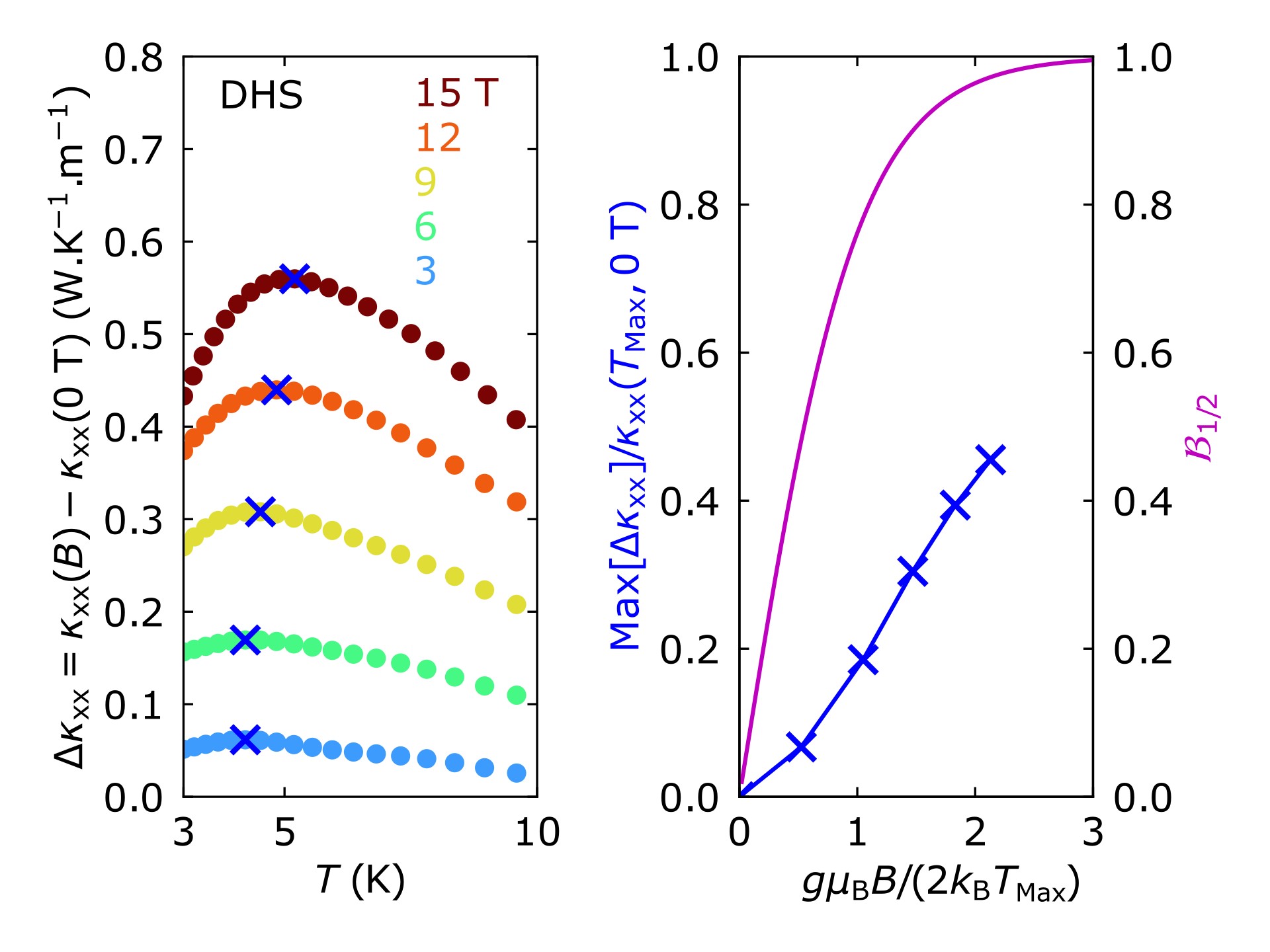}
\caption{\label{fig:Publi-Figure-6}(a) For the DHS sample, difference between finite field and zero field data as a plot of $\Delta\kappa_{\mathrm{xx}}=\kappa_{\mathrm{xx}}(B)-\kappa_{\mathrm{xx}}(0~\mathrm{T})$ vs $T$. The maxima positions $[T_{\mathrm{Max}},\mathrm{Max}(\Delta\kappa_{\mathrm{xx}})]$ are indicated with blue crosses. (b) Left axis (blue): maximal relative increase as a plot of $\mathrm{Max}(\Delta\kappa_{\mathrm{xx}})/\kappa_{\mathrm{xx}}(T_{\mathrm{Max}},0~\mathrm{T})$ vs $g\mu_{\mathrm{B}}B/(2k_{\mathrm{B}}T_{\mathrm{Max}})$, taking $g=2.2$. Right axis (magenta): normalized magnetization $\mathcal{B}_{1/2}$ of paramagnetic (fully free) spins $1/2$, for comparison. Both trends are not found compatible.}
\end{figure}

From the data, we extract the field evolution of $-\xi$ at several fixed temperatures between $3$ and $10$~K so that we can fit $\lambda^{\mathrm{kago}}$, $\lambda^{\mathrm{def}}$ and $m^{\mathrm{def}}(B)$, for which we use a Hill function with two parameters ($u$ and $v$): $m^{\mathrm{def}}(B)=B^{v}/(u^{v}+B^{v})$. It turns out to be impossible to fit $m^{\mathrm{def}}$ with a Brillouin function, hence the choice of this more generic form. Here, $m^{\mathrm{kago}}(T,B)$ is set to an upper bound, based on the analysis presented in the specific heat study~\cite{Barthelemy2022}. For the temperatures and fields under scrutiny, it is temperature-independent, proportional to the magnetic field and reaches values only slightly above $5$~\% in $15$~T: $m^{\mathrm{kago}}(B)\simeq0.0035B$. The fits are very good with this four-parameter model ($\lambda^{\mathrm{kago}}$, $\lambda^{\mathrm{def}}$, $u$, $v$) and the deduced temperature evolutions for the two $\lambda$ couplings reveal that the scattering by magnetic defects becomes dominant below about $8$~K, see Fig.~\ref{fig:Publi-Figure-5}. Besides, we observe that the deduced Hill functions depicting $m^{\mathrm{def}}$ do not resemble the typical Brillouin functions that are routinely used when treating the magnetic defects as nearly free spins $1/2$. They have an initial convex increase (instead of linear) and saturate more slowly, see Fig.~\ref{fig:Publi-Figure-5}. Independent of this fit analysis, the same trend is visible in the evolution of $\mathrm{Max}[\kappa_{\mathrm{xx}}(T,B)-\kappa_{\mathrm{xx}}(T,B=0)]/\kappa_{\mathrm{xx}}(T_{\mathrm{Max}},B=0)$ as a function of $g\mu_{\mathrm{B}}B/(2k_{\mathrm{B}}T_{\mathrm{Max}})$ (taking $g=2.2$), see Fig.~\ref{fig:Publi-Figure-6}. This key result may help to better understand the nature of some of the magnetic defects, at least those that behave as the main scatterers. Indeed, the $m^{\mathrm{def}}$ curves we extract are most likely linear combinations of several magnetization curves, each of them characterizing a specific type of defect that scatters the phonons to a different extent. As $\lambda^{\mathrm{def}}$ is shared, each coefficient in these linear combinations would incorporate a specific defect concentration and coupling renormalization. A complete model, which would account for at least two types of defects as suggested by ESR and NMR studies~\cite{Zorko2017,Khuntia2020}, would be over-parameterized. Our simple model is not designed to obtain quantitative information on $m^{\mathrm{def}}$, although it is already convincing because we note that we retrieve the typical values for the total magnetization when combining about $20$~\% of $m^{\mathrm{def}}$ and $80$~\% of $m^{\mathrm{kago}}$, on the order of $15$~\% for $T=5$~K and $B=15$~T~\cite{Bert2007}. Then, while considering the magnitudes and the nonmonotonic temperature evolution of the $m^{\mathrm{def}}$ curves, see Fig.~\ref{fig:Publi-Figure-5}, we have to keep in mind that $m^{\mathrm{kago}}$ may be overestimated and that we do not separate all types of magnetic defects. Our approach is primarily qualitative and the initial convex increase of the $m^{\mathrm{def}}$ curves is robust.

\subsection{Very low temperature regime ($T<2$~K)}

\begin{figure}[t!]\centering
\includegraphics[width=\linewidth]{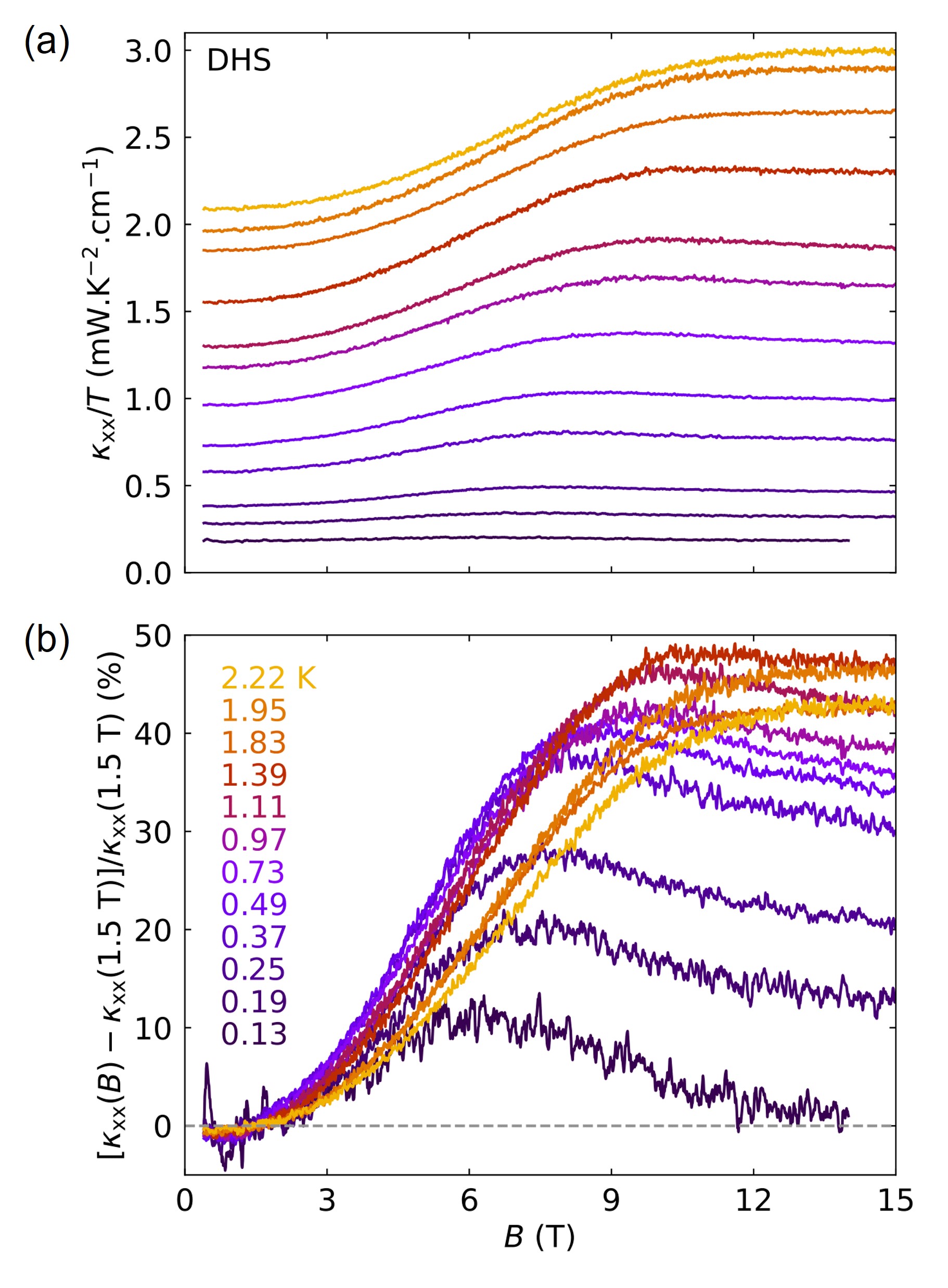}
\caption{\label{fig:Publi-Figure-7}(a) Thermal conductivity of the DHS sample, as a plot of $\kappa_{\mathrm{xx}}/T$ vs $B$ for various temperatures between $0.13$ and $2.22$~K. (b) Relative variations as a plot of $[\kappa_{\mathrm{xx}}(B)-\kappa_{\mathrm{xx}}(1.5\mathrm{~T})]/\kappa_{\mathrm{xx}}(1.5\mathrm{~T})$ vs $B$. The color scale is the same in (a) and (b).}
\end{figure}

The field dependence is less prominent at the lowest temperatures, see Fig.~\ref{fig:Publi-Figure-2}. In our data, it remains visible down to about $0.13$~K and is weaker for the PHS sample than for the DHS sample. It is even weaker in previously reported measurements: it is barely noticeable in the data by Huang \emph{et al.} ($0$, $3$, $6$, $12$ and $15$~T data up to $0.8$~K) and invisible in the data by Murayama \emph{et al.} ($0$ and $14$~T data up to $0.5$~K)~\cite{Huang2021,Murayama2021}. These differences observed from one sample to another are consistent with an essentially extrinsic origin for the field dependence over the whole temperature range, as pointed out in Sec.~\ref{sec:A}: the DHS sample contains more magnetic defects than the PHS sample. On this basis, the sample used by Huang \emph{et al.} seems to have a slightly lower magnetic defect concentration than the PHS sample. The data by Murayama \emph{et al.} are insufficient to make a guess about the magnetic defect concentration in their sample. Let us stress that the absolute value of the thermal conductivity depends on (i) the uncertainty on the geometrical factors and (ii) the densities of all the scatterers (spin excitations in the kagome planes, structural defects, etc.). In this regard, it is not surprising that the thermal conductivity is higher for the DHS sample, see Fig.~\ref{fig:Publi-Figure-2}: if the concentration of magnetic defects is higher, there are fewer spin excitations in the kagome planes to scatter the phonons. 

To provide a detailed view of the peculiar field dependence in the very low temperature regime, we show some field scans at fixed temperatures obtained for the DHS sample in the DR below $2.22$~K, see Fig.~\ref{fig:Publi-Figure-7}(a). Here, it becomes non monotonic below about $1.8$~K, which is made very clear when examining the relative variations $[\kappa_{\mathrm{xx}}(B)-\kappa_{\mathrm{xx}}(1.5\mathrm{~T})]/\kappa_{\mathrm{xx}}(1.5\mathrm{~T})$, see Fig.~\ref{fig:Publi-Figure-7}(b). In this non monotonic regime, the thermal conductivity is found to be almost constant up to about $1.5$~T, then increases until it reaches a maximal value for fields of about $6-11$~T (the maximum seems closer to $6$~T in the data by Huang \emph{et al.}) and eventually decreases moderately as the field intensity continues to rise. It seems that a saturation value would be reached if the field intensity was pushed even further.

For temperatures below about $2$~K, $^{17}$O NMR, torque, susceptibility, inelastic neutron scattering and specific heat measurements have already shown the existence of a characteristic temperature scale defining the very low temperature regime of herbertsmithite~\cite{Helton2009PhD,Jeong2011,Asaba2014,Han2016,Khuntia2020,Barthelemy2022}. This temperature scale is most likely determined by the deviations from the pure nearest neighbor Heisenberg model (e.g. three-dimensional couplings through interplane magnetic defects and their extensions into the kagome planes, Dzyaloshinskii-Moriya interactions, etc.). It features a field-induced transition from the QSL regime to a random frozen phase with gapped excitations and a field-induced crossover of unclear origin~\cite{Jeong2011,Barthelemy2022}. The transition occurs for magnetic fields higher than $1.5$~T at sub-Kelvin temperatures and is not detected in our experiments. The crossover occurs for magnetic fields higher than $10$~T and seems to correspond to the inflection point of $[\kappa_{\mathrm{xx}}(B)-\kappa_{\mathrm{xx}}(1.5\mathrm{~T})]/\kappa_{\mathrm{xx}}(1.5\mathrm{~T})$ at high fields, before saturation. 

The initial increase of $[\kappa_{\mathrm{xx}}(B)-\kappa_{\mathrm{xx}}(1.5\mathrm{~T})]/\kappa_{\mathrm{xx}}(1.5\mathrm{~T})$ for fields up to $6-11$~T is observed whatever the temperature below $2.22$~K and indicates that the spins are progressively polarized, whether we are referring to two decoupled spin systems in the QSL regime (the magnetic defects and the intrinsic spins in the kagome planes) or the whole ensemble of frozen spins in the glassy regime at the lowest temperatures. Below about $1.8$~K, the decrease of the thermal conductivity at higher fields (namely the appearance of non monotonic field dependence) reveals that new scatterers appear. It is then natural to think of spin excitations resulting from a partially polarized state, like ferromagnetic magnons, to scatter the phonons. These spin excitations are expected to vanish at sufficiently high fields so that the phonon thermal conductivity should level off at the value given by $\hat{\kappa}_{\mathrm{xx}}^{\mathrm{phon}}$ and then reach the boundary scattering limit at the lowest temperatures. Yet, down to $0.13$~K, field scans extrapolate to lower values than $\hat{\kappa}_{\mathrm{xx}}^{\mathrm{phon}}$ at high fields, suggesting that a substantial amount of spin excitations remains active.

\section{Discussion}

\subsection{Phonons as the main carriers of heat}

To summarize, with the absence of a residual term in the $T\rightarrow0$ limit in zero field, of visible decoupling at low temperatures and of a sizeable thermal Hall effect, we have shown that there is no clear sign pointing to a direct contribution from neutral fermions to the thermal conductivity. Phonons remain the main carriers of heat from the highest down to the lowest temperatures. This does not mean that the emergent spin excitations are immobile, as they may simply be constrained to delocalize over limited portions of the sample. Still, the absence of spin mediated thermal transport calls into question the class of the ground state. In particular, one must ask whether the seemingly local character of the emergent spin excitations is an intrinsic property of the pure Heinsenberg model or rather a consequence of perturbations such as magnetic defects.

On the theory side, it seems now necessary to test the robustness of the leading QSL proposals to the presence of magnetic defects and determine how the mobility of the spinons is impacted. Various studies already focused on the local response (which may be screened by a spinon Kondo effect~\cite{Florens2006,Gomilsek2019}) expected around vacancies and spinless/spinfull defects in several types of QSL ground states~\cite{Kolezhuk2006,Gregor2008,Rousochatzakis2009,Patil2020}, but did not really examine how the full QSL response is modified for a modest defect concentration as in the case of herbertsmithite. Only the case of a large defect concentration inducing a severe quenched bond randomness was treated and found to result in the stabilization of a gapless valence bond glass made of random singlets~\cite{Singh2010,Kawamura2014,Kimchi2018}. Adapting this scenario to the case of herbertsmithite requires an unrealistic distribution of couplings $\Delta J/J$ on the order of $0.4$~\cite{Kawamura2014}. The upper bound set by $^{17}$O NMR is at least an order of magnitude smaller~\cite{Khuntia2020}.

For some time now, the $Z_{2}[0,\pi]$ ($\alpha$ or $\beta$) and parent $U(1)[0,\pi]$ gapless QSL have been regularly proposed as leading candidates for the ground state of the pure Heisenberg model~\cite{Ran2007,Hermele2008,Clark2013,Iqbal2013,Iqbal2014,He2017,Lu2017,Zhu2018,Hering2019,Zhang2020}. They are constructed from the slave boson approach and involve spinons with a nodal or very small circular Fermi surface. The field independent specific heat $C_{\mathrm{p}}^{\mathrm{kago}}$ tends to rule them out in herbertsmithite, as we expect a field-induced response in presence of weakly interacting fermionic excitations. Let us emphasize however that the spinons are not completely free to move even when deconfined. In particular, the mean field ansatz of a $Z_{2}$ QSL, similar to the Bogoliubov-de-Gennes Hamiltonian of a superconductor, includes a spinon pairing term~\cite{Lu2017}. This term, as it further complexifies the interactions between spinons on the lattice, leads to fascinating properties and may help to explain the lack of a significant spinon contribution to the thermal conductivity, as well as other experimental results. First, very much like Cooper pairs in a superconductor, it is possible that spinon pairs do not carry entropy and thus heat. Second, the nearest neighbor spin correlations become anisotropic so that the establishment of the QSL breaks some lattice symmetries, causing the unpaired spinons to move only along the directions of the most correlated bonds~\cite{Clark2013}. In this vein, ESR, optical ellipsometry and wavelength-dependent multi-harmonic optical polarimetry studies revealed a subtle monoclinic distortion in herbertsmithite, that appears to grow with the building up of short range correlations~\cite{Zorko2017,Laurita2019}. If this slight structural change is really driven by the formation of the QSL, it is then very likely that the spinons in herbertsmithite move in specific directions. The most correlated bonds do not necessarily form connected paths, especially if several domains nucleate or if some magnetic defects split some segments of the paths. Such a scenario would naturally account for the lack of spin-mediated thermal transport.

\subsection{New characteristic temperature scale}

The temperature of about $20$~K below which the thermal conductivity becomes clearly field dependent and deviates from the Debye-Callaway limit in the absence of any magnetic scattering ($\hat{\kappa}_{\mathrm{xx}}^{\mathrm{phon}}$) appears to be a new characteristic temperature for herbertsmithite. In the QSL regime, we attributed the small amplitude and the monotonic field dependence of the thermal conductivity to a strong scattering of phonons from emergent excitations of the QSL and from magnetic defects. The occurrence of these two scattering processes seems simultaneous at all fields and points to the establishment of the QSL regime. Indeed, as long as the temperature is high enough to keep all the spins in a conventional paramagnetic regime, the spins that will be eventually involved in the magnetic defects and in the QSL regime at lower temperatures remain indistinguishable. The magnetic defects appear only when the QSL settles, entangling some spins and excluding others.

In fact, we notice that this temperature of about $20$~K coincides with the temperature at which multiple contributions (one ``main'' and two ``defect'' sets of $5$ peaks) begin to clearly appear in the $^{17}$O NMR spectra~\cite{Khuntia2020}. Correspondingly, the spin-lattice relaxation rate determined with $^{17}$O NMR and Cu NQR starts to combine a fast component and a slow component~\cite{Khuntia2020,Wang2021}. However, it does not seem suitable to connect this temperature with other special temperatures already identified in herbertsmithite. First, torque measurements revealed that a change of magnetization easy axis occurs at lower temperatures on the order of $12$~K~\cite{Zorko2017}. Indeed, below this threshold, the susceptibility of the magnetic defects considerably exceeds the susceptibility of the spins involved in the QSL regime. There is no reason for this change of anisotropy to yield the onset of the field dependence and we further note that our data do not present any anomaly in the relevant temperature range to signal this behavior. Second, NMR measurements showed that the $^{35}$Cl spin-lattice relaxation rate peaks at much higher temperatures on the order of $50$~K, simultaneously with a sudden change of the $^{17}$O quadrupolar frequency at some defect sites~\cite{Imai2008,Olariu2008,Fu2015}. This was attributed to the freezing of some OH bonds with random orientations. Again, there is no clue for this subtle structural transition in our data, nor obvious relation with the onset of field dependence. 

Then, thanks to the strong spin-lattice coupling in herbertsmithite, it appears that thermal transport is a tool of choice to detect the subtle crossover towards the QSL regime. At the end, magnetic defects are both problematic and useful. On the one hand, the QSL class cannot be properly characterized as soon as a substantial amount of magnetic defects is likely to affect the mobility of the emergent excitations: fully localized spin excitations or barely mobile spin excitations lead to the same absence of spin mediated heat transport. On the other hand, the onset of field dependence combined with the appearance of two magnetic scattering processes implies that magnetic defects have nucleated on top of a new underlying state which encompasses all other spins. In this regard, magnetic defects help to reveal the establishment of the QSL. These observations call for complementary measurements on other QSL candidate materials for which we could vary the concentration of magnetic defects, as in the Zn-brochantite, Zn-barlowite, Zn-claringbullite and Zn-averievite families by adjusting the zinc content.

\subsection{Negligible phonon Hall effect}

We did not detect any sizeable thermal Hall effect in herbertsmithite. On the one hand, this is an indication that the spin excitations, if mobile, are bad heat carriers with respect to phonons. Indeed, deconfined spinons that delocalize with a large mean free path should give rise to a significant Lorentz force driven thermal Hall effect, which is expected to scale with their direct contribution to the longitudinal conductivity (null here)~\cite{Katsura2010}. On the other hand, it is clear that there is no sign of a phonon thermal Hall effect either. We put these observations in perspective with results obtained on other insulators. 

Four frustrated magnets with a kagome geometry have already been studied through thermal Hall effect measurements, namely Cu($1-3$,bdc), volborthite, Ca-kapellasite and Cd-kapellasite~\cite{Hirschberger2015,Watanabe2016,Doki2018,Akazawa2020}. In practice, these materials have fewer magnetic defects than herbertsmithite but involve competing anisotropic interactions, resulting in very different magnetic models. Unlike herbertsmithite, they eventually order at low temperature. In all four cases, there is evidence that some mobile spin excitations contribute to the heat conduction along with the phonons and generate a sizeable thermal Hall effect: magnons in Cu($1-3$,bdc) and bosonic spinons in volborthite, Ca-kapellasite and Cd-kapellasite. In particular, the thermal Hall effect of volborthite, Ca-kapellasite and Cd-kapellasite is very well captured by the Schwinger-boson mean-field theory from the slave fermion approach.

There are other evidences for a thermal Hall effect from mobile spin excitations in various non kagome materials. Actually, the first magnon thermal Hall effect was observed in the pyrochlore ferromagnet Lu$_{2}$V$_{2}$O$_{7}$~\cite{Onose2010}. Later on, the thermal Hall effect detected in the quantum spin ice pyrochlore Tb$_{2}$Ti$_{2}$O$_{7}$ was assigned to more exotic spin excitations~\cite{Hirschberger2015-bis}. More recently, the Kitaev spin liquid candidate $\alpha$-RuCl$_{3}$ came under the spotlight with several conflicting measurements of a thermal Hall effect, either half-quantized and ascribed to Majorana fermions, or not quantized and ascribed to bosonic edge excitations such as topological magnons~\cite{Yokoi2021,Bruin2022,Czajka2022}. 

It is now established that phonons can also generate a thermal Hall effect, whether in presence or absence of magnetic ions. This effect was first observed in $2005$ in the paramagnetic garnet Tb$_{3}$Ga$_{5}$O$_{12}$~\cite{Strohm2005,Inyushkin2007}. More recently, it was observed in a wide variety of systems, often with an unexpectedly large magnitude: the multiferroic ferrimagnetic Fe$_{2}$Mo$_{3}$O$_{8}$~\cite{Ideue2017}, some cuprates~\cite{Grissonnanche2019,Boulanger2020,Grissonnanche2020}, the quantum paraelectric SrTiO$_{3}$~\cite{Li2020}, the cubic antiferromagnet Cu$_{3}$TeO$_{6}$~\cite{Chen2022}, the kagome antiferromagnet Cd-Kapellasite~\cite{Akazawa2020} and the Kitaev QSL candidate $\alpha$-RuCl$_{3}$~\cite{Lefrancois2022}. Yet, the phonons are not chiral particles and, despite growing theoretical developments on that matter, the reasons why they can be preferentially deflected in one direction or the other depending on the sign of the applied magnetic field still remain enigmatic. 

Among the proposed scenarios, there are some intrinsic mechanisms, where the vibrational modes strongly couple to the ionic environment and its defining electric and magnetic moments. These moments shall interact with the applied magnetic field, sometimes in a very subtle way only, and this may eventually give rise to a detectable time reversal odd contribution to the Berry curvature of phonons~\cite{Qin2012}. We emphasize that intrinsinc mechanisms are generally predicted to result in a small amplitude, that would become significant under very special circumstances only. In the case of SrTiO$_{3}$ for instance, an intrinsic effect may arise due to the flexoelectric coupling of phonons to their nearly ferroelectric environment, but is predicted to be orders of magnitude smaller than the actual observation~\cite{Li2020,Chen2020}. However, if they do not average to zero because of domains, some intrinsic mechanisms could be argued in the case of Fe$_{2}$Mo$_{3}$O$_{8}$, and possibly the cases of Cd-Kapellasite and $\alpha$-RuCl$_{3}$, where a strong spin lattice coupling may prevail~\cite{Ideue2017,Kocsis2022}.

There are also some extrinsic mechanisms, where time reversal oddness may originate from skew scattering and side jump from defects~\cite{Guo2021,Sun2022,Guo2022,Flebus2022}. Extrinsinc mechanisms are generally predicted to result in a larger magnitude and have been invoked to explain the observations on Tb$_{3}$Ga$_{5}$O$_{12}$ and SrTiO$_{3}$. In Tb$_{3}$Ga$_{5}$O$_{12}$, the phonon Hall effect is attributed to a resonant scattering from the crystal field states of superstoichiometric Tb$^{3+}$ defects~\cite{Mori2014}. In SrTiO$_{3}$, it is attributed to a scattering from twin boundaries between tetragonal domains~\cite{Li2020,Chen2020}.

In principle, herbertsmithite has both the intrinsic and extrinsic ingredients to enable the generation of a phonon Hall effect, namely a strong spin-lattice coupling and a substantial number of extrinsic scatterers. Yet, the observed signal is vanishingly small in both the paramagnetic and QSL regimes, albeit the magnitude of the longitudinal thermal conductivity is higher than that detected in $\alpha$-RuCl$_{3}$ and Cd-Kapellasite~\cite{Akazawa2020,Lefrancois2022}. It underlines that the phonon Hall effect occurs in a magnetic insulator only when specific requirements are met for (i) the nature of the intrinsic magnetic state (ordered or disordered, frozen or dynamical) and (ii) the type of the defects (magnetic or non magnetic), so that the spin textures developing around the defects drive the phonons in a preferred direction depending on the sign of the magnetic field.

\section{Conclusion}

We have measured the thermal conductivity and thermal Hall conductivity of herbertsmithite. Given the absence of a residual term in the $T\rightarrow0$ limit in zero field, of visible decoupling at low temperatures and of a sizeable thermal Hall effect, we conclude that deconfined spinons, if present, are not contributing significantly to the heat conduction in the QSL regime. Phonons are found to be the main carriers of heat down to the lowest temperatures. Our data reveal that the thermal conductivity, which is field-independent at high temperatures, acquires a clear field dependence below about $20$~K -- monotonic above about $2$~K and nonmonotonic below. The onset of the monotonic field dependence is mainly ascribed to the polarization of some magnetic defects and is thought to signal the establishment of the QSL regime. By extracting the magnetization profile of some of the magnetic defects, we show that they are not isolated paramagnetic objects. This calls for further investigations to understand their spatial extension and their impact on the kagome planes. The nonmonotonic field dependence is found compatible with previously reported instabilities at very low temperatures and suggests that magnetic fields larger than $15$~T are required to fully polarize the field-induced glassy phase. \\

\underline{Note added:} While finalizing this Article, it came to our attention that Murayama \emph{et al.} have recently supplemented their data~\cite{Murayama2022}. Still restricted to magnetic fields of $0$ and $14$~T, their data now range from $0.1$ to $5$~K ($0.1$ to $0.5$~K before~\cite{Murayama2021}). In contrast to our findings, they report an identical thermal conductivity for both fields, up to $5$~K, and conclude that the spin-phonon coupling is vanishingly small.

\acknowledgments

We thank B. Fauqué, L. Messio and J. A. Quilliam for helpful discussions. We would like to acknowledge the support of the French Agence Nationale de la Recherche, under Grant No. ANR-18-CE30-0022 «LINK». V. Balédent acknowledges the MORPHEUS platform at the Laboratoire de Physique des Solides. L. Taillefer acknowledges support from the Canadian Institute for Advanced Research (CIFAR) as a CIFAR Fellow and funding from the Institut Quantique, the Natural Sciences and Engineering Research Council of Canada (PIN: 123817), the Fonds de Recherche du Québec -- Nature et Technologies, the Canada Foundation for Innovation, and a Canada Research Chair. This research was undertaken thanks in part to funding from the Canada First Research Excellence Fund.

\end{document}